# REGISTRATION OF THERMAL NEUTRONS USING TWO-DETECTOR MODULE BASED ON SILICON DETECTORS AND METALLIC GADOLINIUM

*O.S. Deiev, V.N. Dubina, S.K. Kiprich, V.D. Ovchinnik, G.P. Vasiliev, V. I. Yalovenko*
*National Science Center "Kharkov Institute of Physics and Technology", Kharkov, Ukraine*

The thermal neutrons were registered using a two-detector module based on two planar Si detectors: with a neutron converter from natural metallic Gd and without a converter. To obtain thermal neutrons we used a $^{239}$Pu-Be($\alpha$,n) fast neutron source with a paraffin moderator. The measurements were carried out simultaneously by two detectors by processing two channels of registration with a special computer program. In experiments gamma radiation was not suppressed. The conversion electrons of the nuclear reaction Gd(n,$\gamma$e)Gd were registered. We performed the registration of thermal neutrons with the possibility of subtraction for background radiation.
PACS: 28.20.Gd, 07.05.Tp

## 1. INTRODUCTION

The prospectivity of thermal neutrons registration using semiconductor Si detector and a converter from Gd were shown in the works [1-3]. Gadolinium has the largest thermal neutron capture cross section (up to 300,000 barns).

This work continues the research of recent years at the NSC KIPT on registration thermal neutrons using planar Si detectors with a converter of neutrons from natural metallic Gd [4-10]. In GEANT4 we were simulated the slowing down of neutrons and the yield of the nuclear thermal-neutron capture reaction Gd(n,$\gamma$e) Gd [4,5]. Sealed modules for detecting radiation were developed [6, 7] and the peculiarities of radiation registration by Si pin detectors were studied [8, 9]. In work [10] the conversion electrons from capture reaction and the CXR Gd were measured experimentally by a Gd-Si detector system in a single-channel mode. The background gamma radiation was measured by a Si detector without a Gd layer. The measurements were carried out sequentially in equal geometry. The gamma radiation of the neutron source $^{239}$Pu-Be($\alpha$,n) was cut off by a protective layer of lead and iron. Thus, the thermal neutron flux was registered in the absence of gamma background.

The purpose of this work is the creation and testing of a two-detector module based on two uncooled silicon detectors and metallic gadolinium for detection of thermal neutrons with the possibility of subtracting background gamma radiation. The measurements were carried out simultaneously by two detecting channels using a special program. In experiments gamma radiation was not suppressed.

The method was based on the separation of the conversion electrons (CE) from gadolinium converter in mixed spectra from gamma-electron flux.

## 2. MATERIALS AND METHODS

Natural gadolinium contains two important isotopes: $^{155}$Gd-14.8% (cross-section 17 000 barn), $^{157}$Gd - 15.7% (cross-section 300 000 barn). Neutron capture reactions with a maximum cross section:

n + $^{155}$Gd → $^{156}$Gd$^*$ → $^{156}$Gd + $\gamma$-quanta + Gd CXR + conversion e$^-$ (39-199 keV);

n + $^{157}$Gd → $^{158}$Gd$^*$ → $^{158}$Gd + $\gamma$-quanta + Gd CXR + conversion e$^-$ (29-182 keV).

In this work the CXR Gd lines with energies $K_\alpha$ = 42.99 keV, $K_\beta$ = 48.69 keV and conversion electrons in the energy range 30...200 keV were registered.

To obtain thermal neutrons, we used a fast neutron source $^{239}$Pu-Be($\alpha$,n) with a paraffin retarder. Neutron sources emit neutrons and gamma rays in a wide range of energies. The intensity of gamma radiation is usually comparable or noticeably higher than the neutron flux.

A gadolinium foil 0.4 mm thick was placed close to the Si detector (Fig.1). The design of the thermal neutron registration module is described in detail in [6].

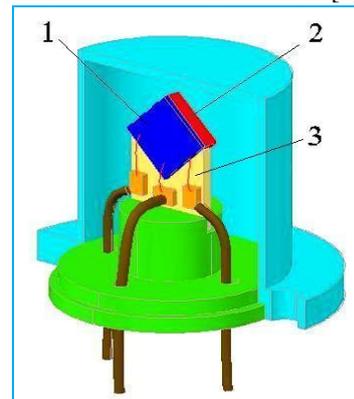

*Fig. 1. General view of the sealed module: 1 – Si detector, 2 – Gd converter, 3 – sital substrate*

To calibrate spectrometric systems and create mixed electronic and gamma radiation fields radioactive sources $^{241}$Am, $^{57}$Co, $^{90}$Sr-$^{90}$Y, $^{137}$Cs were used.

In Fig. 2 shows the scheme of the experiment used in this work. Two-detector system that realizes the separation of the gamma-neutron flux **is** shown. In the two-detector module there are two uncooled Si pin detectors with area 5x5 mm$^2$ and thickness of 300 μm. One of the placed detectors has a Gd converter. The inverse geometry is applied when the Gd layer is located behind the Si detector. D is the thickness of paraffin between the source of fast neutrons and the detector. The thickness of paraffin (D, cm) can vary from 0 to 10 cm in increments of 2 cm. The Al plate 0.5 mm thick is part of the design of the sealed detector. The experiments did not use protective foils of Pb, Fe to cut off background gamma radiation.

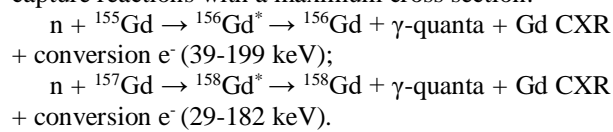

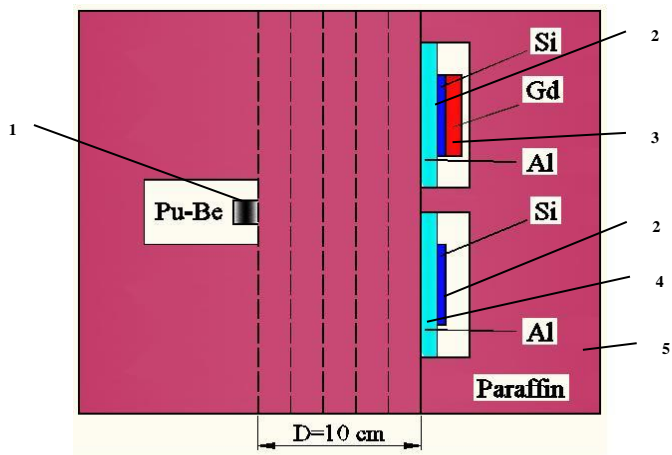

*Fig. 2. Schematic of the experiment in geometry, when Gd is located behind the Si detector: 1 – $^{239}$Pu-Be (α,n) neutron source, 2 – Si detector, 3 – Gd converter, 4 – Al case, 5 – paraffin moderator*

Assuming equality of the gamma responses of both detectors, the distribution obtained by the detector without Gd foil, corresponding to the background gamma response, is subtracted from the spectral distribution of the first detector with Gd converter. Thus a useful signal of the capture reaction Gd(n, γe)Gd was separated.

In Fig. 3 shows a two-detector module for registration thermal neutrons.

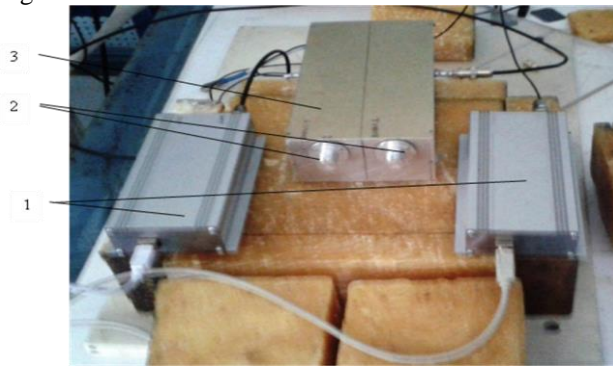

*Fig. 3. Two-detector module for recording thermal neutrons: 1 – analog-to-digital converters, 2 –detectors Si-Gd and Si, 3 – spectrometric electronics*

The module uses two uncooled silicon detectors with similar characteristics in separate cases. The detector designs are similar to each other, but unlike the second detector, the first detector has a Gd converter. There are two spectrometric amplifiers in the center. Then paraffin was placed on all sides. The preamplifier of the spectrometric tract is located inside the paraffin. The neutron source is placed on the vertical axis in the middle between the two detecting modules. This provides the equal gamma and neutron radiation for the two detectors.

A program for simultaneous reading of signals from two detectors was developed. The window of the program and the spectrum set are shown in Fig. 4.

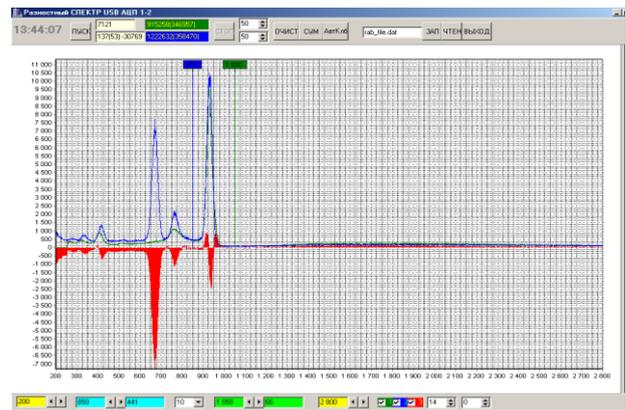

*Fig. 4. The window of the program for dialing the spectrum from the source $^{241}$Am. The blue line is the detector with Gd, the green line is the detector without Gd, the red line is the spectrum difference*

Simultaneous measurement of spectra by two detectors were performed. The radiation lines from the $^{241}$Am source are shown in the Fig. 4. Also shown the "difference" (subtraction) of two spectra - red line.

Fig. 5 shows the location of the two-detector system and source in the paraffin moderator. Fig. 5, a - a source of neutrons placed close to the detectors. Fig. 5, b - the process of placing the detector in paraffin.

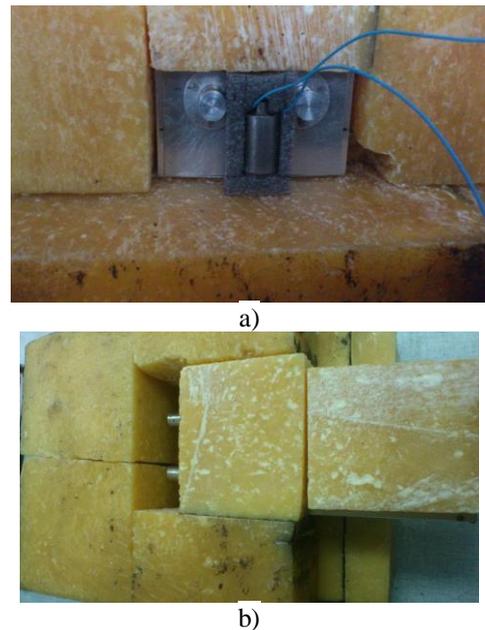

a)

b)

*Fig. 5. The locations of two-detector system and source in paraffin moderator*

### 3. RESULTS AND DISCUSSION

Some possible difficulties in performing measurements and deductions of two spectra obtained by two spectrometric systems based on Si planar detectors with a Gd converter and without a converter are shown in Fig. 6.

In Fig. 6a, the calculated radiation peaks with energies at a maximum of 59 and 60 keV are presented. The widths at half-height FWHM are 1.5 keV. In Fig. 6, b shows the calculated radiation peaks with energies at a maximum of 60 keV. The widths at half-height FWHM are 1.7 and 1.5 keV. The results of subtraction of spectra are also shown.

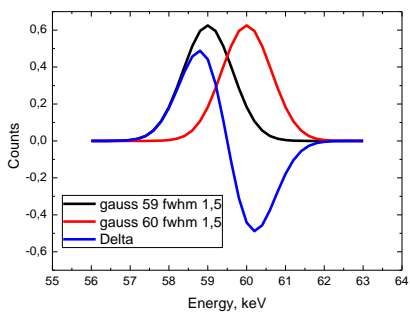

*Fig. 6, a. Calculated radiation peaks with energies at a maximum of 59 and 60 keV. The widths at half height of the FWHM are 1.5 keV. Delta is the result of subtraction of two spectra*

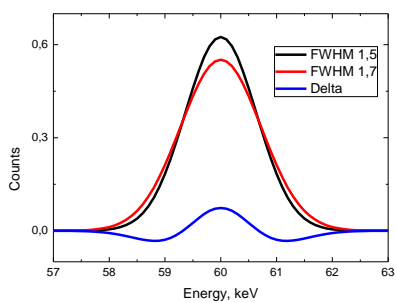

*Fig. 6, b. Calculated emission peaks with energies at a maximum of 60 keV. The widths at half height FWHM are 1.7 and 1.5 keV. Delta is the result of subtracting spectra*

As can be seen, even a slight difference in the position of the radiation peaks associated with the different amplification of the spectrometric channel, or the widths of the radiation peaks associated with the detector resolution, will lead to a characteristic distortion of the difference (subtraction) of the spectra. That is, instead of the expected zero level of the signal, one can get peaks in the residue spectrum.

Let us go to the experiments on recording thermal neutrons by two detectors, and to identification of the conversion electrons of the Gd (n,γe) Gd reaction. No protective foils in experiments were used to cut off gamma radiation.

The responses of the two detectors to gamma radiation of $^{239}$Pu-Be (α,n) of the fast neutron source were preliminarily measured. The paraffin moderator was completely removed from all sides. The response of the detector was formed by fast neutrons and gamma radiation from source.

In Fig. 7 shows the experimental spectra measured by two identical detecting modules Si, 5x5 mm² (with a gadolinium converter and without a Gd converter). The spectrum Delta is the difference between two radiation spectra.

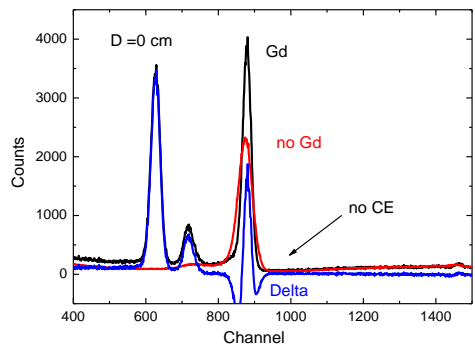

*Fig. 7. Experimental spectra measured by two detection modules Si, 5×5 mm² with gadolinium converter and without Gd converter. Moderator and protection removed. Delta is the difference between two radiation spectra. No conversion electrons (CE) observed*

As can be seen, the resolution of the detectors is slightly different in this two measurements. Therefore when we subtracted the spectra the distortion of the Delta spectrum arises in the energy region of 59.54 keV. This distortion is shown in Fig.6, b.

The line with enrgy 59.54 keV, CXR peaks from Gd and a small background from a neutron source are observed. No distinct structures are observed after the 900 channel (above 60 keV) in the spectra and in the Delta spectrum is not observed. In subsequent experiments we used detectors with the same energy resolution FWHM ~ 1.55 keV.

The calculated spectra of conversion electrons in the neutron capture in the natural Gd are presented in Fig. 8 [4,10].

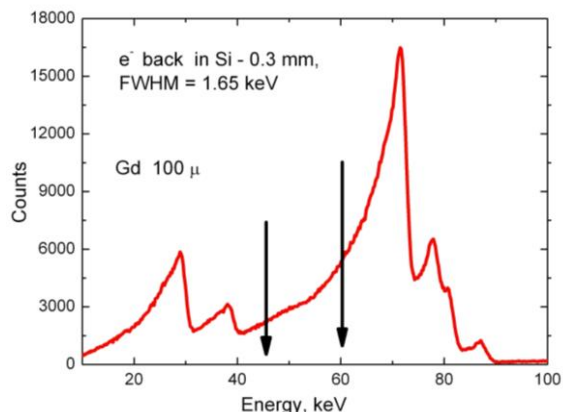

*Fig. 8. Calculated spectra of conversion electrons in the neutron capture reaction in natural Gd*

Two groups of conversion electrons are observed in the form of a spectrum with two maxima with energies of 70 and 80 keV. The arrows show the positions of the CXR Gd lines and gamma-ray lines with an energy of 59.54 keV.

Fig. 9 shows experimental spectra measured by two detecting modules Si, 5x5 mm2 with Gd converter and without Gd. A layer of paraffin D = 2 cm was used.

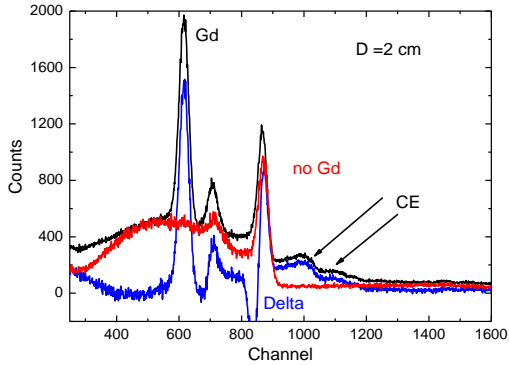
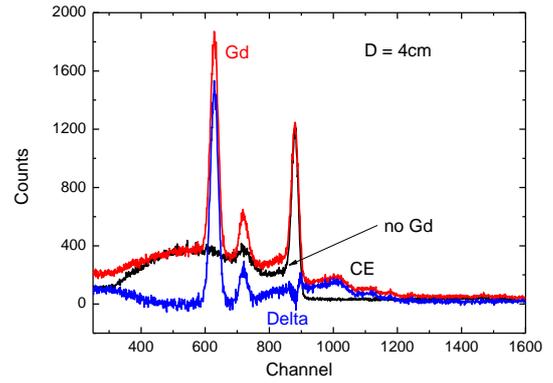

*Fig. 9. Experimental spectra measured by two detection modules Si, 5×5 mm2 with gadolinium converter and without Gd. Delta is the difference between the radiation spectra. D = 2 cm*

*Fig. 10. Experimental spectra measured by two detection modules Si, 5×5 mm2 with gadolinium converter and without Gd. Delta is the difference between the radiation spectra. D = 4 cm*

The two-maximum structure in the form of a wide distribution in the region > 60 keV (> 900 channel) was observed. This is the spectrum of conversion electrons (CE), corresponding to groups of electrons with energies of ~ 70 and 80 keV. The fine structure of the conversion electron spectrum is not visible, which is due to the resolution of the silicon detector FWHM ~ 1.55 keV, as well as to the straggling of the electron energies as they pass through the gadolinium layer and the protective passivating Si layer of the detector.

Thus, the experimental spectra contain two groups of conversion electrons (CE), CXR Gd lines in the presence of a Gd converter, a gamma-ray line with an energy of 59.54 keV from the $^{239}$Pu-Be (α, n) neutron source. As can be seen, it was possible experimentally to register conversion electrons from the neutron capture reaction Gd (n,γe) Gd.

In addition, the spectrum has a characteristic wide maximum in the range of 300-800 channels. Here gamma quanta of reverse and multiple Compton scattering from the 59.54-keV line are recorded. Thus an additional wide maximum is formed in the experimental spectrum.

We can note that even in the case of incomplete compensation of the peak at 59.54 keV, the group of conversion electrons is clearly distinguished. It is possible to compare the number of conversion electrons with the calculated data and determine the averaged value of the thermal neutron flux.

The result of the experimental measurement for D = 4 cm is shown in Fig. 10. Two experimental spectra are measured by two detecting modules Si, 5x5 mm² with a gadolinium converter and without a Gd converter.

As can be seen, the conversion electrons have practically cleared from the background gamma radiation. There remains a slight distortion of the spectrum, associated with an incorrect subtraction of the 59.54-keV line. A group of conversion electrons stands out clearly. Here it is possible to approximate the spectrum of conversion electrons in the range of 750-1000 channels.

Similar measurements were made for D = 10 cm. The results of the experimental measurements are shown in Fig. 11. Two experimental spectra are measured by two detecting modules Si, 5x5 mm² with a Gd converter and without a Gd converter.

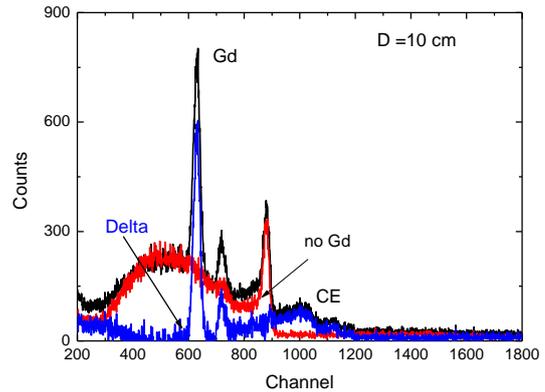

*Fig. 11. Experimental spectra measured by two detection modules Si, 5×5 mm² with gadolinium converter and without Gd. Delta is the difference between the radiation spectra. D = 10 cm*

In this experiment, the distortions of the Delta spectrum (difference) are minimal. Conversion electrons are clearly distinguished.

For the paraffin thickness D = 4 cm, an experiment was carried out with two successive measurements. We performed two measurements with a gadolinium converter, differing from each other by rotating the detector around the axis by 180 ° (inverted position). Thus, we measured the yield of the nuclear neutron capture reaction in the front and back geometries. The results of the experimental measurement for D = 4 cm are shown in Fig. 12. Two experimental spectra are measured in front and back geometry, and also without a converter.

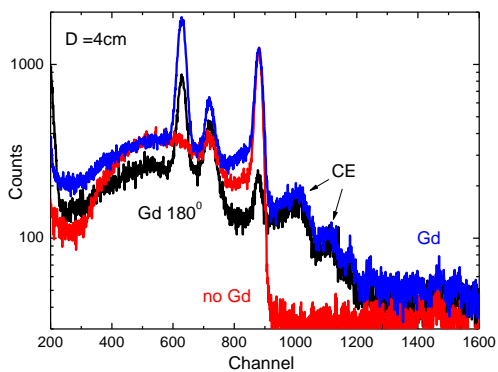

*Fig. 12. Two experimental spectra are measured in front (blue) and back (black) geometry, and also without a converter (red). D = 4 cm*

Two groups of conversion electrons are practically equal. This measurement shows that the flow of thermal neutrons from the source in the geometry chosen by us is practically isotropic. In the process of multiple scattering, the neutron flux is redistributed and within the paraffin cube the direction of their motion is isotropic.

**CONCLUSION**

A small-size thermal neutron detection module based on two Si detectors and a metal Gd converter was developed and tested. Detectors were used with an area of 5x5 mm$^2$ and a thickness of 0.3 mm. To obtain thermal neutrons, we used a $^{239}$Pu-Be ($\alpha$, n) fast neutron source with a paraffin moderator.

The experimental spectra for the Gd (n, γe) Gd reaction were measured. The spectra consisted of Gd CXR lines with energy of 42.99 and 48.69 keV and conversion electrons in the energy range 30 ... 200 keV with a maximum at energy of ~ 70 keV. The background gamma radiation was measured by a Si detector without a Gd layer. Protection from radiation was not applied. The measurements were carried out simultaneously by two detectors by processing two channels of registration with a special computer program.

Studies have shown that the use of a two-detector module allows thermal neutrons to be registered using the yield of conversion electrons of the Gd(n,γe)Gd reaction. We performed the registration of thermal neutrons with the possibility of subtraction for background radiation.

Publications are based on the research provided by the grant support of the State Fund for Fundamental Research (project F79/39197). The study was conducted within the IDEATE of the International Associated Laboratory (LIA) [11].